\documentclass[11pt,a4paper]{article}
\usepackage[left=15mm, top=10mm, right=15mm, bottom=20mm, nohead=true, nofoot=true]{geometry}

\usepackage[utf8]{inputenc}
\usepackage{authblk}
\usepackage{indentfirst}
\usepackage{misccorr}
\usepackage{graphicx}
\usepackage{amsmath}
\usepackage{amssymb}
\usepackage{multicol}
\usepackage{bm}
\usepackage{longtable}
\usepackage{pdflscape}
\usepackage{epstopdf}
\usepackage{multirow}
\usepackage{adjustbox}
\usepackage{amsfonts}
\usepackage{ifthen}
\usepackage{fancyhdr}
\usepackage{setspace}
\usepackage{xcolor}
\usepackage[round,authoryear,comma,sort&compress,numbers]{natbib}
\usepackage[toc,title,page]{appendix}
\usepackage[hidelinks]{hyperref}
\usepackage{url}
\usepackage{caption}

\newcommand{\msun}{\mbox{$\,{\rm M}_\odot$}}
\newcommand{\lsun}{\mbox{$\,{\rm L}_\odot$}}

\hypersetup{
    colorlinks=true,
    linkcolor=green,
    citecolor=blue,
    filecolor=magenta,
    urlcolor=cyan,
    pdftitle={Sharelatex Example},
    pdfpagemode=FullScreen,
}

\fancypagestyle{plain}{\chead{\texttt{\textcolor{brown}{Communications of BAO, Vol. ?, Issue ?, 20??, pp. ???-???}}}}
\title{\textbf{SEDust: A Pipeline for Deriving Best-Fit Spectral Energy Distributions from DUSTY Outputs}}
\author[1]{Hamidreza Mahani \thanks{mahani@ipm.ir}}
\author[1]{Atefeh Javadi}
\affil[1]{\scriptsize School of Astronomy, Institute for Research in Fundamental Sciences (IPM), P.O. Box 1956836613, Tehran, Iran}
\def\apj{Astrophys.~J. }

\def\aap{Astron.~Astrophys. }

\def\apjs{Astrophys.~J.~Suppl.~Ser. }
\def\aj{Astron.~J. }
\def\mnras{Mon.~Not.~R.~Astron.~Soc. }

\def\aapr{Astron.~Astrophys.~Rev. }

\begin{document}
\pagestyle{empty}
\newpage
\pagestyle{fancy}
\label{firstpage}
\date{}
\maketitle

\begin{abstract}

In this study, we introduce \texttt{SEDust}, a pipeline designed to identify the best-fitting spectral energy distributions from the outputs of the DUSTY code and compare them to observational data. The pipeline incorporates a grid of 24000 models, enabling robust fitting for both carbon- and oxygen-rich AGB stars. It calculates key physical parameters, including luminosity, optical depth, and mass-loss rate, and produces the corresponding best-fit SED plots. Using \texttt{SEDust}, we derived the specific mass-return rates for the galaxies NGC 147 and NGC 185. The specific mass-return rate of AGB stars in NGC 147 is $8.13 \times 10^{-12}  \mathrm{yr}^{-1}$, while in NGC 185 it is $6.52 \times 10^{-11}  \mathrm{yr}^{-1}$. These results indicate that the mass loss from evolved stars alone cannot account for the total mass budget required to sustain these galaxies, highlighting the need for additional sources or mechanisms of mass replenishment to resolve the observed discrepancies.

\end{abstract}


\section{Introduction}

The mass loss and dust production of evolved stars, particularly those on the asymptotic giant branch, play a key role in the dynamical evolution and chemical enrichment of galaxies by contributing material to the interstellar medium \citep{Boyer25, Abdollahi23, Javadi11}. Understanding the processes that govern dust and star formation and their dependence on stellar and environmental properties is therefore essential for insights into galactic evolution \citep{Mahani23}. Dwarf elliptical (dE) galaxies, such as NGC 147 and NGC 185, which are satellites of Andromeda, offer valuable laboratories for studying these phenomena and are of considerable interest in modern observational astronomy because they provide important insights into the interpretation of deep redshift surveys and the evolution of stars and interstellar medium particles \citep{Ferguson94}. While both galaxies share similarities in luminosity \citep{Crnojevic14} and velocity dispersion \citep{Geha10}, they exhibit notable differences. 

NGC 147 contains no gas or dust and shows no evidence of recent star formation, whereas NGC 185 hosts interstellar material and shows signs of recent star formation in its central regions \citep{Young97,Welch98,Marleau10,DeLooze16}. Photometry of the red giant branch indicates that NGC 147 has a slightly higher mean metallicity than NGC 185 \citep{Davidge94,McConnachie05,Geha10}. Their proximity on the sky (58 arc-min) has led some to suggest they form a bound galaxy pair \citep{vandenBergh98}, though kinematic studies indicate they may not be gravitationally bound \citep{Geha10,Watkins13}. Additionally, NGC 147 shows signs of tidal distortion, whereas NGC 185 does not \citep{Ferguson16}. Differences in star formation histories further suggest that the amounts of gas, mass loss, and dust in their interstellar media differ \citep{Hamedani17, Hamedani_Golshan17_II}. In this study, we aim to quantitatively test this hypothesis by estimating the mass-loss and dust injection rates in these galaxies.

\section{Data}

To investigate the role of evolved stars in shaping the interstellar medium of NGC 147 and NGC 185, we compiled multi-wavelength datasets from several well-established catalogs. These include surveys of long-period variables, carbon-star populations, and dust-enshrouded AGBs obtained through optical, near-infrared, and mid-infrared observations. The combination of these catalogs provides a comprehensive view of the late stages of stellar evolution in both galaxies, capturing variability, chemical composition, and dust production. Such an approach enables a detailed assessment of mass-loss processes and their dependence on stellar type and environment, which is particularly relevant given the contrasting star formation histories, and interstellar medium contents of the two galaxies discussed in the introduction \citep{Hamedani17, Mahani25}.

\subsection{Catalogs}
\label{sec:catalogs}

Long-period variables were cataloged by \citet{Lorenz11} using i-band time-series photometry from the ALFOSC instrument on the Nordic Optical Telescope over 2.5 years with more than 30 epochs, supplemented by single-epoch $K_s$-band and narrow-band data. They identified 213 LPVs in NGC 147 and 513 in NGC 185, with periods between 90 and 800 days and amplitudes up to 2 mag. Near-infrared surveys of carbon stars were conducted by \citet{Sohn06} and \citet{Kang05} using CFHTIR, identifying 91 stars in NGC 147 and 73 in NGC 185 with J, H, K photometry. The DUSTiNGS survey \citep{Boyer15a,Boyer15b,Boyer17} observed both galaxies at 3.6 and 4.5 µm across two epochs to detect variability and heavily dust-enshrouded extreme AGB stars (x-AGBs), estimating roughly 124 in NGC 147 and 99 in NGC 185. Photometric uncertainties in these datasets range from 0.02 to 0.27 mag.

Optical observations by \citet{Nowotny03, Nowotny01} employed V, i, TiO, and CN filters to classify O-rich (M-type) and C-rich (C-type) stars in fields of $6.5' \times 6.5'$, identifying 146 new C-stars in NGC 147 and 154 in NGC 185. \citet{Battinelli147,Battinelli185} further expanded the carbon-star census using the CN–TiO technique with the CFH12K camera, finding 145 C-stars in NGC 185 with a mean absolute magnitude of –4.41 ± 0.05 mag. Together, these multi-band observations provide a robust foundation for constructing spectral energy distributions and studying mass-loss rates of evolved stars in these dwarf elliptical galaxies.

\subsection{SEDs}
\label{sec:code}

By cross-matching the available catalogs, we reconstructed the spectral energy distributions (SEDs) of LPV stars in NGC 147 and NGC 185 and incorporated their atmospheric chemistry into the modeling. The chemical classification of asymptotic giant branch (AGB) stars into oxygen-rich (C/O $<$ 1) and carbon-rich (C/O $>$ 1) types is a critical factor, since their atmospheric composition strongly affects the emergent SEDs \citep{Nowotny03}. This distinction arises from the third dredge-up process, which enriches the stellar envelope with carbon, and is particularly efficient in metal-poor environments where carbon stars form more readily \citep{Mowlavi99}. Properly accounting for this chemistry is therefore essential when comparing theoretical SEDs to observations.

To model these systems, we employed the radiative transfer code \texttt{DUSTY} \citep{Ivezic97}, which simulates the interaction of radiation with circumstellar dust in a spherically symmetric geometry. The code solves the radiative transfer equation while incorporating absorption, scattering, and thermal re-emission, enabling realistic SEDs of dust-enshrouded stars \citep{Javadi13_III, Javadi22, Abdollahi23_II}. Its flexibility in handling a wide range of dust grain properties makes it particularly suited for AGB stars with diverse dust environments. Beyond circumstellar shells, \texttt{DUSTY} has been widely applied to star-forming regions, interstellar dust, and active galactic nuclei, demonstrating its versatility in deriving physical parameters such as dust temperature, density distributions, and stellar mass-loss rates from observed spectra.

\begin{figure}
\begin{center}
\includegraphics[width=0.8\textwidth]{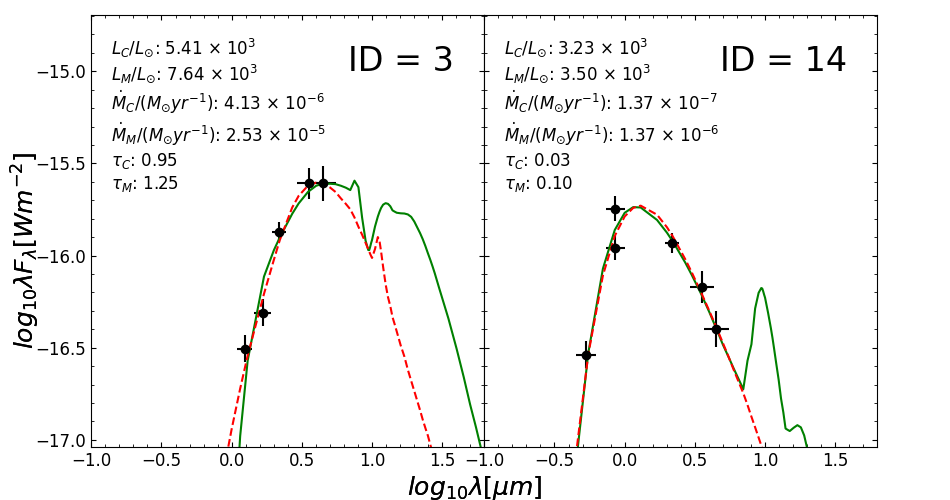}
\caption{ \texttt{SEDust} results for two example stars from the target galaxies (left: NGC 147, right: NGC 185). Stellar IDs are adopted from \citet{Mahani25}. The solid green curve represents the best-fit oxygen-rich model, while the red dashed curve corresponds to the best-fit carbon-rich model. Photometric uncertainties for each band are shown as error bars, based on the values reported in the original catalogs.}
\label{fig:2x2}
\end{center}
\end{figure}

Given the large number of LPVs and AGB stars in our sample, manually running \texttt{DUSTY} through trial-and-error fitting is not feasible. To address this, we developed \texttt{SEDust}\footnote{\url{https://github.com/hmahani/SEDust}}, an automated pipeline that employs a $\chi^{2}$ minimization scheme to determine the best-fitting SEDs by comparing \texttt{DUSTY} model kernels with observational data. \texttt{SEDust} performs a systematic grid-based parameter exploration, incorporating variations in stellar temperature (2000–3000 K, in 100 K steps), dust temperature (500–1500 K, in 100 K steps), and optical depth (0.01–2.0, with fine increments of 0.01 below 0.5 and 0.05 above 0.5). The chemical composition of the dust was tailored to the stellar type: carbon stars were modeled with a mixture of 85\% amorphous carbon \citep{Hanner88} and 15\% silicon carbide \citep{Pegourie88}, while oxygen-rich stars were modeled with astronomical silicates \citep{Draine84}. An illustration of the pipeline’s performance for different stars in both galaxies is presented in Figure \ref{fig:2x2}.

\texttt{SEDust} was executed independently for oxygen-rich and carbon-rich AGB stars, enabling the pipeline to account for intrinsic chemical differences. The resulting grid comprised approximately 12,000 models for each stellar type, ensuring a robust coverage of parameter space. This approach allowed us to efficiently and consistently capture the diversity of SEDs across both oxygen- and carbon-rich populations, providing a reliable framework for deriving stellar and dust properties.

\section{Results}

Utilizing the photometric catalogs described in Section \ref{sec:catalogs}, and employing the \texttt{SEDust} pipeline outlined in Section \ref{sec:code}, we derived the physical properties of AGB stars in the environments of NGC 147 and NGC 185. The use of \texttt{SEDust} allowed us to fit the SEDs of large samples of LPVs and AGB stars in a fully automated manner, ensuring consistency across the datasets. Through this method, we obtained estimates of luminosity, mass-loss rates, and dust injection rates, which serve as crucial diagnostics of the contribution of evolved stars to the interstellar medium in these two dwarf elliptical galaxies.

For NGC 147, we find that the luminosity of AGB stars spans the range from $(6.1 \pm 0.25) \times 10^{2} \lsun$ to $(7.8 \pm 0.32) \times 10^{3} \lsun$. In contrast, stars in NGC 185 show a broader luminosity distribution, extending from $(5.7 \pm 0.23) \times 10^{2} \lsun$ up to $(1.6 \pm 0.07) \times 10^{4} \lsun$. This difference suggests that NGC 185 currently hosts a larger population of bright, dust-producing AGB stars, consistent with its younger stellar population and evidence of more recent star formation compared to NGC 147 \citep{Hamedani17}. The derived luminosities provide not only a measure of the energy output of evolved stars but also a key input parameter for constraining their evolutionary status \citep{Gholami25}.

In terms of mass-loss, we estimate that the contribution from LPV stars amounts to $(7.8 \pm 3.1) \times 10^{-5}\ \msun\,\mathrm{yr}^{-1}$ in NGC 147 and $(3.0 \pm 1.2) \times 10^{-4}\ \msun\,\mathrm{yr}^{-1}$ in NGC 185. When considering all AGB stars in the sample, the total mass-loss rates increase to $(9.4 \pm 3.8) \times 10^{-4}\ \msun\,\mathrm{yr}^{-1}$ for NGC 147 and $(1.6 \pm 0.6) \times 10^{-3}\ \msun\,\mathrm{yr}^{-1}$ for NGC 185. These values imply that AGB stars in NGC 185 are returning significantly more material to the interstellar medium than those in NGC 147, highlighting once again the difference in their stellar content and evolutionary histories. The corresponding dust mass injection rates are $(5.9 \pm 2.4) \times 10^{-6}\ \msun\,\mathrm{yr}^{-1}$ for NGC 147 and $(9.9 \pm 4.0) \times 10^{-6}\ \msun\,\mathrm{yr}^{-1}$ for NGC 185, further underlining the enhanced role of NGC 185 in enriching its surrounding medium with dust.

\begin{figure}
\begin{center}
\includegraphics[width=0.49\textwidth]{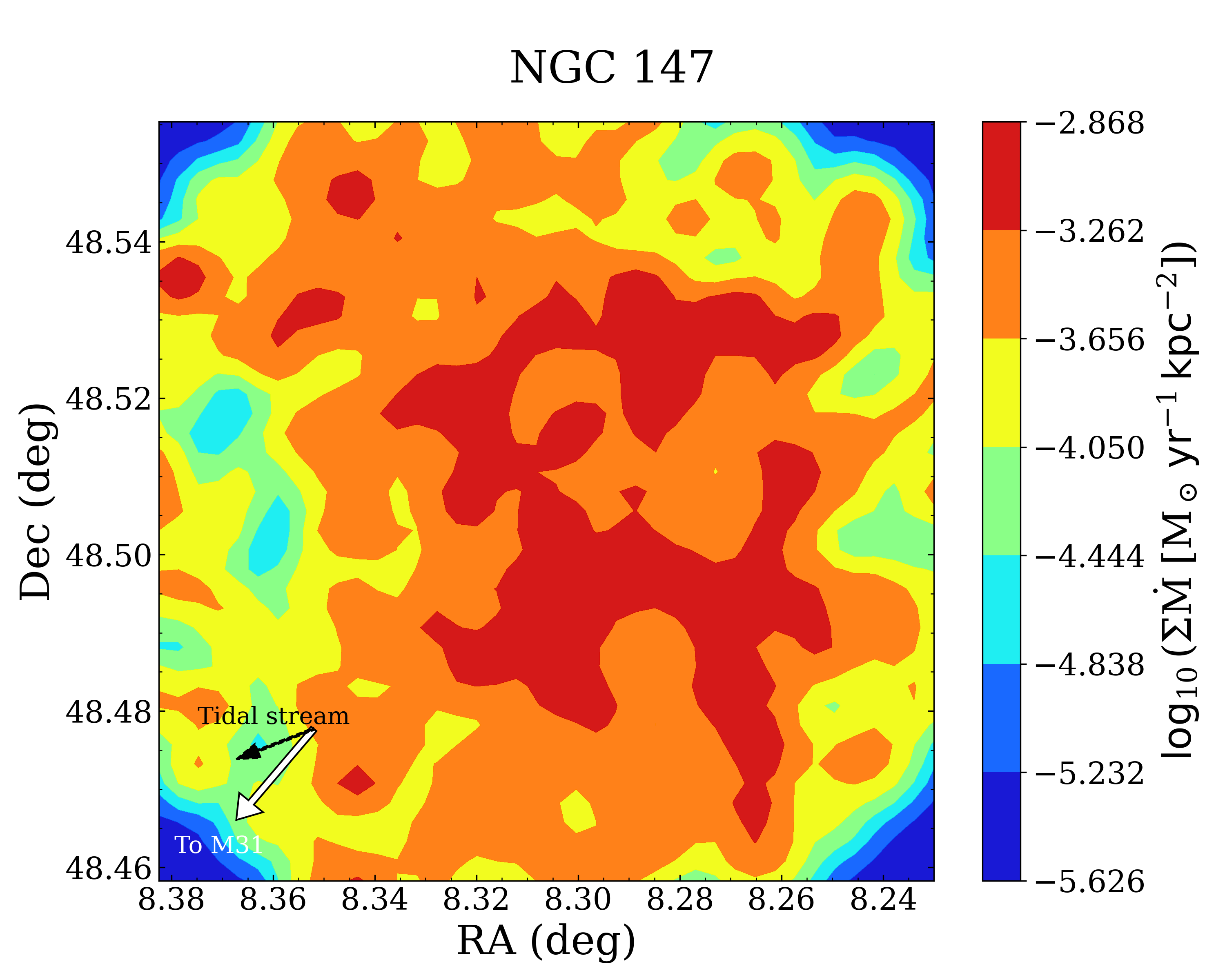}
\includegraphics[width=0.49\textwidth]{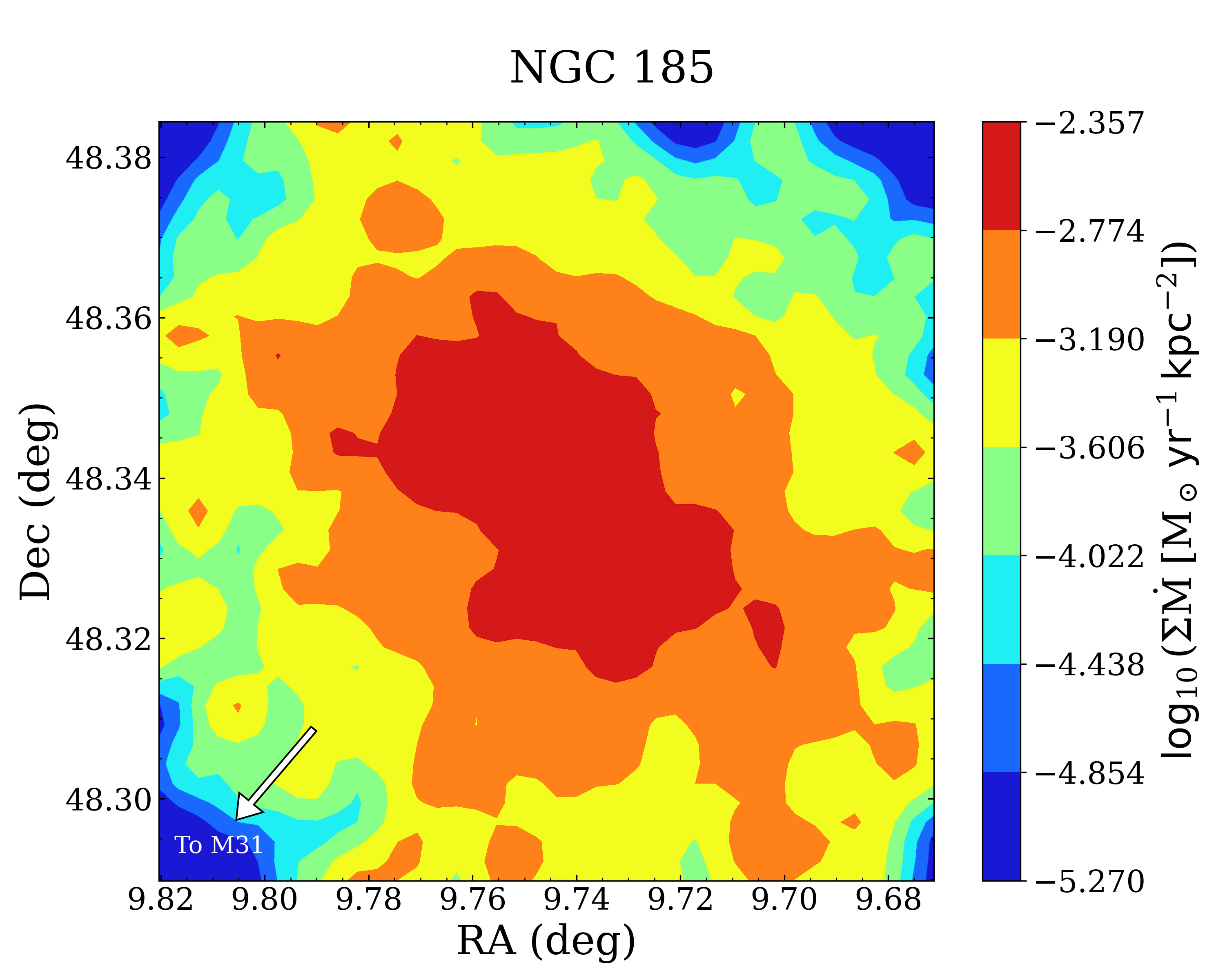}
\caption{Surface density maps of AGB star mass-loss rates in NGC 147 and NGC 185. The white arrows mark the approximate direction toward M31, while the black arrow in the NGC 147 panel highlights the orientation of the tidal stream.}
\label{fig:Mdot_Contour}
\end{center}
\end{figure}

Figure \ref{fig:Mdot_Contour},  a two-dimensional map of dust dispersion, indicates that gravitational interactions between NGC 147 and the Andromeda galaxy are likely influencing the distribution of dust within NGC 147. These external forces may be responsible for disturbing or stripping its interstellar material, contributing to the observed lack of recent star formation in this galaxy \citep{Hamedani17}. In contrast, NGC 185, which appears less affected by such interactions, retains a larger reservoir of gas and dust and continues to show signs of relatively recent star-forming activity. This environmental distinction between the two galaxies provides additional context for interpreting the measured differences in luminosity and mass-loss properties of their evolved stars.

The results of our analysis are summarized in Table~\ref{tab:results_summary}, which presents the luminosity ranges, mass-loss rates, dust injection rates, and specific mass-return rates for both galaxies. The table also lists the replenishment timescales implied by these mass-return rates, which significantly exceed the Hubble time in both cases. This discrepancy highlights that AGB stars alone cannot supply sufficient material to sustain the stellar populations of these galaxies, and that additional sources of gas and dust (such as high-mass stars, supernovae, or external accretion) are required to explain their observed properties.

\begin{table}[ht]
\centering
\caption{Summary of derived properties for AGB stars in NGC 147 and NGC 185.}
\label{tab:results_summary}
\begin{tabular}{lcc}
\hline
\hline
\textbf{Parameter} & \textbf{NGC 147} & \textbf{NGC 185} \\
\hline
Luminosity range [$\lsun$] & $6.1 \times 10^{2}$ -- $7.8 \times 10^{3}$ & $5.7 \times 10^{2}$ -- $1.6 \times 10^{4}$ \\
LPV mass-loss rate [$\msun\,\mathrm{yr}^{-1}$] & $(7.8 \pm 3.1)\times10^{-5}$ & $(3.0 \pm 1.2)\times10^{-4}$ \\
Total AGB mass-loss rate [$\msun\,\mathrm{yr}^{-1}$] & $(9.4 \pm 3.8)\times10^{-4}$ & $(1.6 \pm 0.6)\times10^{-3}$ \\
Dust injection rate [$\msun\,\mathrm{yr}^{-1}$] & $(5.9 \pm 2.4)\times10^{-6}$ & $(9.9 \pm 4.0)\times10^{-6}$ \\
Specific mass-return rate [$\mathrm{yr}^{-1}$] & $8.13\times10^{-12}$ & $6.52\times10^{-11}$ \\
Mass replenishment timescale [yr] & $1.23\times10^{11}$ & $1.53\times10^{10}$ \\
\hline
\end{tabular}
\end{table}

\section{Conclusion}

\begin{itemize}

\item We developed and applied a new automated pipeline, \texttt{SEDust}, designed to fit  SEDs of AGB stars using the radiative transfer code \texttt{DUSTY}. By exploring a comprehensive parameter grid, \texttt{SEDust} enables efficient and consistent determination of key stellar properties such as luminosity, optical depth, and mass-loss rates for large stellar samples.

\item The pipeline was executed separately for oxygen-rich and carbon-rich AGB stars, with more than 24000 models in total. This approach ensured that chemical composition effects were appropriately accounted for in the modeling, thereby improving the robustness of the derived physical parameters.

\item Applying \texttt{SEDust} to AGB populations in NGC 147 and NGC 185, we measured luminosities, dust temperatures, and mass-loss rates. The total mass-loss rates from AGB stars were estimated to be $(9.4 \pm 3.8) \times 10^{-4} M_\odot \mathrm{yr}^{-1}$ for NGC 147 and $(1.6 \pm 0.6) \times 10^{-3} M_\odot \mathrm{yr}^{-1}$ for NGC 185. These translate into specific mass-return rates of $8.13 \times 10^{-12} \mathrm{yr}^{-1}$ for NGC 147 and $6.52 \times 10^{-11} \mathrm{yr}^{-1}$ for NGC 185.

\item The inverse of these specific rates implies replenishment timescales of $1.23 \times 10^{11} \mathrm{yr}$ for NGC 147 and $1.53 \times 10^{10} \mathrm{yr}$ for NGC 185, both longer than the Hubble time. This demonstrates that the mass returned by evolved stars alone is insufficient to sustain the present-day stellar populations of these galaxies.

\item These results highlight the necessity of additional sources of gas and dust, such as contributions from high-mass stars, supernovae, or external accretion, to explain the observed material content of dwarf galaxies. The \texttt{SEDust} pipeline thus provides a powerful and scalable tool for quantifying stellar feedback and constraining the life cycle of baryonic matter in galaxies.

\end{itemize}

%\section*{\small Acknowledgements}
%\scriptsize{The ComBAO would like to the thank the dedicated researchers who are publishing with the ComBAO.}

\scriptsize
\bibliographystyle{ComBAO}
\nocite{*}
\bibliography{SEDust_ComBAO}

\appendix
\renewcommand{\thesection}{\Alph{section}.\arabic{section}}
\setcounter{section}{0}
\normalsize
\begin{appendices}
\section{Sample of Outputs}

\begin{longtable}{@{}cccccccccccc@{}}
\caption{A sample of results for NGC 147. The ID column lists the unique identifier of each star in the final catalog. Subscripts C and M correspond to the best-fit carbon and oxygen models, respectively. $\tau$ is the optical depth, $L$ the luminosity, and $\dot{M}$ the mass-loss rate. $BC$ denotes the bolometric correction, and $M_{ini}$ the initial stellar mass estimated as described in \citep{Mahani25}. The “Chemical type” column gives the chemical classification reported in the literature, while the “Mass type” column provides the classification inferred from the initial mass.} \label{tab:147results} \\
\hline
\hline
\textbf{ID} & \boldmath$\tau_{C}$ & \boldmath$\tau_{M}$ & \boldmath$L_{C}$ & \boldmath$L_{M}$ & \boldmath$\dot{M}_{C}$ & \boldmath$\dot{M}_{M}$ & \boldmath$BC_{C}$ & \boldmath$BC_{M}$ & \boldmath$M_{ini}$ & \textbf{Chemical type} & \textbf{Mass type} \\
 &  &  & [\lsun] & [\lsun]& [\msun/yr] & [\msun/yr] & [mag] & [mag]& [\msun] & &  \\
\hline
\endfirsthead

\multicolumn{12}{c}%
{{\tablename\ \thetable{} -- continued from previous page}} \\
\hline
\hline
\textbf{ID} & \boldmath$\tau_{C}$ & \boldmath$\tau_{M}$ & \boldmath$L_{C}$ & \boldmath$L_{M}$ & \boldmath$\dot{M}_{C}$ & \boldmath$\dot{M}_{M}$ & \boldmath$BC_{C}$ & \boldmath$BC_{M}$ & \boldmath$M_{ini}$ & \textbf{Chemical type} & \textbf{Mass type} \\
 &  &  & [\lsun] & [\lsun]& [\msun/yr] & [\msun/yr] & [mag] & [mag]& [\msun] & &  \\
\hline
\endhead

\hline \multicolumn{12}{|r|}{{Continued on next page}} \\ \hline
\endfoot

\hline
\endlastfoot

1	&	0.01	&	0.01	&	3550	&	3400	&	2.07E-08	&	1.87E-07	&	3.23	&	3.28	&	1.07	&	C	&	M	\\
2	&	0.2	&	0.41	&	3510	&	4130	&	1.17E-06	&	5.66E-06	&	3.17	&	3	&	1.11	&	C	&	C	\\
3	&	0.95	&	1.25	&	5410	&	7640	&	4.13E-06	&	2.53E-05	&	2.52	&	2.14	&	1.01	&	C	&	M	\\
4	&	0.01	&	0.01	&	6220	&	6420	&	3.03E-08	&	3.01E-07	&	3.15	&	3.12	&	1.73	&	C	&	C	\\
5	&	0.02	&	0.02	&	3370	&	3390	&	3.36E-08	&	1.3E-07	&	3.25	&	3.24	&	1.18	&	C	&	C	\\
6	&	0.06	&	0.01	&	3180	&	3170	&	9.65E-08	&	5.74E-08	&	3.33	&	3.34	&	1.15	&	C	&	C	\\
7	&	0.85	&	1.15	&	4540	&	5650	&	2.16E-06	&	1.31E-05	&	3.05	&	2.81	&	1.31	&	C	&	C	\\
8	&	0.01	&	0.01	&	4770	&	4940	&	2.58E-08	&	2.62E-07	&	3.08	&	3.04	&	1.35	&	C	&	C	\\
9	&	0.02	&	0.01	&	5710	&	6380	&	5.69E-08	&	3E-07	&	3.14	&	3.02	&	1.63	&	C	&	C	\\
10	&	0.01	&	0.01	&	7420	&	7560	&	3.33E-08	&	3.21E-07	&	3.09	&	3.07	&	2.01	&	C	&	C	\\
11	&	0.01	&	0.01	&	4010	&	4200	&	2.51E-08	&	2.74E-07	&	3.08	&	3.03	&	1.05	&	C	&	M	\\
12	&	0.01	&	0.01	&	4720	&	4780	&	2.56E-08	&	2.56E-07	&	3.19	&	3.18	&	1.34	&	C	&	C	\\
13	&	0.01	&	0.01	&	1660	&	1800	&	1.48E-08	&	1.83E-07	&	2.96	&	2.88	&	0.83	&	C	&	M	\\
14	&	0.02	&	0.01	&	3490	&	3410	&	3.93E-08	&	1.77E-07	&	3.21	&	3.23	&	1.1	&	C	&	C	\\
15	&	0.7	&	0.95	&	4860	&	6520	&	2.67E-06	&	1.62E-05	&	2.74	&	2.42	&	1.19	&	C	&	C	\\

\hline

\end{longtable}

\end{appendices}

\end{document}